\documentclass[12pt]{article}
\usepackage[dvips]{epsfig}
\begin{document}
\begin{flushright}
OHSTPY-HEP-T-98-027 \\
hep-th/9811083
\end{flushright}
\vspace{20mm}
\begin{center}
{\LARGE Super Yang-Mills at Weak, Intermediate and Strong Coupling}
\\
\vspace{20mm}
{\bf Francesco Antonuccio, Oleg Lunin, Stephen Pinsky} \\
\vspace{4mm}
{\em Department of Physics,\\ The Ohio State University,\\ Columbus,
OH 43210, USA}

\end{center}
\vspace{10mm}

\begin{abstract}
We consider three dimensional SU($N$) ${\cal N}=1$ super-Yang-Mills
compactified on the space-time ${\bf R} \times S^1 \times S^1$. In
particular, we compactify the light-cone coordinate
$x^-$ on a light-like circle via DLCQ, and wrap the remaining  transverse
coordinate $x^{\perp}$ on a spatial circle.   By retaining only the first few
excited modes in the transverse direction, we are able to solve for bound
state wave functions and masses numerically by diagonalizing the
discretized light-cone supercharge. This regularization of the theory is
shown to preserve supersymmetry. We plot bound state masses as a
function of the coupling, showing the transition in particle masses
as we move from a weakly to a strongly-coupled theory. 
We analyze both numerically and analytically massless states
which exist only in the limit of strong or weak gauge coupling.
In addition, we find massless
states that persist for all values of the gauge coupling.  An analytical
treatment of these massless states is provided.  Interestingly, in the
strong coupling limit, these massless states become string-like.  

\end{abstract}
\newpage

\def\be{\begin{equation}}
\def\bea{\begin{eqnarray}}
\def\ee{\end{equation}}
\def\eea{\end{eqnarray}}
\def\d{\partial}

\baselineskip .25in

\section{Introduction}
An outstanding challenge in quantum field theory is
solving non-Abelian gauge theories at intermediate and
strong coupling. Recently, there has been considerable
progress in understanding the properties of strongly coupled
gauge theories with supersymmetry
\cite{seibergwitten,seiberg,maldacena}. In particular, there
are a number of supersymmetric gauge theories that are believed
to be inter-connected through a web of strong-weak coupling dualities.
Although existing evidence for
these dualities is encouraging, there is still an urgent need
to address these issues at a more fundamental level. Ideally,
we would like to solve for the bound states of these
theories directly, and at any coupling.

Of course, solving a field theory from first principles
is typically an intractable task. Nevertheless, it has been
known for some time that $1+1$ dimensional field theories
{\em can} be solved from first principles via a straightforward
application of
DLCQ  (see \cite{bpp98} for a review).
In more recent times, a large class of supersymmetric
gauge theories in two dimensions was studied using
a supersymmetric form of DLCQ  (or `SDLCQ'), which is known to
preserve supersymmetry \cite{sakai95,klebhash,alp98a,alp98b,
alpp98,alpp99,alp99}.

Evidently, it would be desirable to extend
these DLCQ/SDLCQ algorithms to solve higher dimensional theories.
One important difference between two dimensional and higher
dimensional theories is the phase diagram induced by
variations in the gauge coupling. The
spectrum of a $1+1$ dimensional gauge theory
scales trivially with
respect to the gauge coupling, while a theory in higher dimensions
has the potential of exhibiting a complex phase structure,
which may include a strong-weak coupling duality.
It is therefore interesting to study
the phase diagram of gauge theories in $D \geq 3$ dimensions.

Towards this end, we consider
three dimensional SU($N$) ${\cal N}=1$ super-Yang-Mills
compactified on the space-time ${\bf R} \times S^1 \times S^1$.
In particular, we compactify the light-cone coordinate
$x^-$ on a light-like circle via DLCQ, and wrap the remaining
transverse coordinate $x^{\perp}$ on a spatial circle.
By retaining only the first few excited modes in the
transverse direction, we are able to solve for bound state
wave functions and masses numerically by diagonalizing the
discretized light-cone
supercharge.
We show that the supersymmetric formulation
of the DLCQ procedure -- which was
studied in the context of two dimensional theories \cite{sakai95,alp99} --
extends naturally in $2+1$ dimensions, resulting in an exactly
supersymmetric spectrum.

The contents of this paper are organized as follows.
In Section \ref{formulation}, we formulate
SU($N$) ${\cal N}=1$ super-Yang-Mills
defined on the compactified space-time ${\bf R} \times S^1 \times S^1$.
Explicit expressions are given for the light-cone supercharges,
which are then discretized via the SDLCQ procedure.
Quantization of the theory is then carried out by imposing
canonical (anti)commutation relations for boson and fermion fields.
In Section \ref{numerical}, we present the results of our
numerical diagonalizations by plotting bound state masses
as a function of gauge coupling.
We also study the bound state
structure of the massless states in the theory.
In Section \ref{analytical}, we provide an analytical treatment
of certain massless states in the theory, and discuss
the appearance of new massless states at strong coupling.
We conclude our analysis with a discussion of our results
in Section \ref{summary}.

\section{Light-Cone Quantization and SDLCQ}
\label{formulation}
We wish to study the bound states of ${\cal N} =1$
super-Yang-Mills in $2+1$ dimensions.
Any numerical approach necessarily involves introducing a
momentum lattice -- i.e.
parton momenta can only take on discretized values.
The usual space--time lattice explicitly breaks supersymmetry,
so if we wish to discretize our theory {\em and} preserve supersymmetry,
then a judicious choice of lattice is required.

In $1+1$ dimensions, it is well known that the light-cone
momentum lattice
induced by the DLCQ procedure
preserves supersymmetry if the supercharge rather than the
Hamiltonian is discretized \cite{sakai95,alp99}.
In $2+1$ dimensions, a supersymmetric prescription
is also possible. We begin by introducing light-cone
coordinates $x^{\pm} = (x^0 \pm x^1)/\sqrt{2}$, and compactifying
the $x^-$ coordinate on a light-like circle.
In this way, the conjugate light-cone momentum $k^+$ is discretized.
To discretize the remaining (transverse) momentum $k^{\perp} = k^2$,
we may compactify $x^{\perp} = x^2$ on a spatial circle.
Of course, there is a significant
difference between the discretized light-cone momenta $k^+$,
and discretized transverse momenta $k_{\perp}$; namely,
the light-cone momentum $k^+$ is always positive\footnote{Since we wish
to consider the decompactified limit in the end, we omit zero
modes. This is a necessary technical constraint in
numerical calculations.}, while $k_{\perp}$ may take on positive
or negative values. The positivity of $k^+$ is a key property
that is exploited in DLCQ calculations;
for any given light-cone compactification,
there are only a finite number of choices for $k^+$ -- the
total number depending on how finely we discretize the
momenta\footnote{The `resolution' of the discretization is
usually characterized by a positive integer $K$, which is called
the `harmonic resolution' \cite{pb85,yamawaki}; for a given choice of $K$,
the light-cone momenta $k^+$ are restricted to positive integer
multiples of $P^+/K$, where $P^+$ is the total light-cone momentum
of a state}.
In the context of two dimensional theories, this implies a finite
number of Fock states \cite{pb85}.

In the case
we are interested in -- in which there is an additional transverse
dimension -- the number of Fock states is no longer finite,
since there are an arbitrarily large number of transverse momentum
modes defined on the transverse spatial circle.
Thus, an additional truncation of the transverse momentum
modes is required to render the total number of Fock states
finite, and the problem numerically tractable\footnote{
This truncation procedure, which is characterized by some
integer upper bound, is analogous to the truncation of $k^+$
imposed by the `harmonic resolution' $K$.}.
In this work, we choose the simplest truncation procedure
beyond retaining the zero mode; namely,
only partons with transverse momentum $k_\perp=0,\pm\frac{2 \pi}{L}$
will be allowed, where $L$ is the size of the transverse circle.

Let us now apply these ideas in the context of a specific
super-Yang-Mills theory.
We start with $2+1$ dimensional ${\cal N}=1$ super-Yang-Mills theory
defined on a space-time with one transverse dimension
compactified on a circle:
\be
S=\int d^2 x \int_0^L dx_\perp \mbox{tr}(-\frac{1}{4}F^{\mu\nu}F_{\mu\nu}+
{\rm i}{\bar\Psi}\gamma^\mu D_\mu\Psi).
\ee
After introducing the light--cone coordinates
$x^\pm=\frac{1}{\sqrt{2}}(x^0\pm x^1)$, decomposing the spinor $\Psi$
in terms of chiral projections --
\be
\psi=\frac{1+\gamma^5}{2^{1/4}}\Psi,\qquad
\chi=\frac{1-\gamma^5}{2^{1/4}}\Psi
\ee
and choosing the light--cone gauge $A^+=0$, the action becomes
\bea\label{action}
S&=&\int dx^+dx^- \int_0^L dx_\perp \mbox{tr}\left[\frac{1}{2}(\d_-A^-)^2+
D_+\phi\d_-\phi+ {\rm i}\psi D_+\psi+ \right.\nonumber \\
& &
\left.
 \hspace{15mm} +{\rm i}\chi\d_-\chi+\frac{{\rm i}}{\sqrt{2}}\psi D_\perp\phi+
\frac{{\rm i}}{\sqrt{2}}\phi D_\perp\psi \right].
\eea
A simplification of the
light--cone gauge is that the
non-dynamical fields $A^-$ and $\chi$ may be explicitly
solved from their Euler-Lagrange equations of motion:
\bea
A^-&=&\frac{g}{\d_-^2}J=
\frac{g}{\d_-^2}\left(i[\phi,\d_-\phi]+2\psi\psi\right),\\
\chi&=&-\frac{1}{\sqrt{2}\d_-}D_\perp\psi.\nonumber
\eea

These expressions may be used to express any operator
in terms of the physical degrees of freedom only.
In particular, the light-cone energy, $P^-$, and momentum
operators, $P^+$,$P^{\perp}$,
corresponding to  translation
invariance in each of the coordinates
$x^\pm$ and $x_\perp$ may be calculated explicitly:
\bea\label{moment}
P^+&=&\int dx^-\int_0^L dx_\perp\mbox{tr}\left[(\d_-\phi)^2+
{\rm i}\psi\d_-\psi\right],\\
P^-&=&\int dx^-\int_0^L dx_\perp\mbox{tr}
\left[-\frac{g^2}{2}J\frac{1}{\d_-^2}J-
    \frac{{\rm i}}{2}D_\perp\psi\frac{1}{\d_-}D_\perp\psi\right],\\
P_\perp &=&\int dx^-\int_0^L dx_\perp\mbox{tr}\left[\d_-\phi\d_\perp\phi+
    {\rm i}\psi\d_\perp\psi\right].
\eea
The light-cone supercharge in this theory
is a two component Majorana spinor, and may be conveniently
decomposed in terms of its chiral projections:
\bea\label{sucharge}
Q^+&=&2^{1/4}\int dx^-\int_0^L dx_\perp\mbox{tr}\left[\phi\d_-\psi-\psi\d_-
           \phi\right],\\
Q^-&=&2^{3/4}\int dx^-\int_0^L dx_\perp\mbox{tr}\left[2\d_\perp\phi\psi+
    g\left({\rm i}[\phi,\d_-\phi]+2\psi\psi\right)\frac{1}{\d_-}\psi\right].
\eea
The action (\ref{action}) gives the following canonical
(anti)commutation relations for
propagating fields at equal $x^+$:
\begin{eqnarray}
\left[\phi_{ij}(x^-,x_\perp),\d_-\phi_{kl}(y^-,y_\perp)\right]
&=&
\frac{1}{2}{\rm i}\delta(x^- -y^-)\delta(x_\perp -y_\perp)
\left( \delta_{il}\delta_{jk} - \frac{1}{N}\delta_{ij}\delta_{kl} \right),
\\
\left\{\psi_{ij}(x^-,x_\perp),\psi_{kl}(y^-,y_\perp)\right\}
&=&
\frac{1}{2}\delta(x^- -y^-)\delta(x_\perp -y_\perp)
\left( \delta_{il}\delta_{jk} - \frac{1}{N}\delta_{ij}\delta_{kl} \right).
\label{comm}
\end{eqnarray}

Using these relations one can check the supersymmetry algebra:
\be
\{Q^+,Q^+\}=2\sqrt{2}P^+,\qquad \{Q^-,Q^-\}=2\sqrt{2}P^-,\qquad
\{Q^+,Q^-\}=-4P_\perp.
\label{superr}
\ee

We will consider only states which have vanishing transverse momentum,
which is possible since the total transverse momentum operator
is kinematical\footnote{Strictly speaking, on a transverse
cylinder, there are separate sectors with total
transverse momenta $2\pi n/L$; we consider only one of them, $n=0$.}.
On such states, the light-cone supercharges
$Q^+$ and $Q^-$ anti-commute with each other, and the supersymmetry algebra
is equivalent to the ${\cal N}=(1,1)$ supersymmetry
of the dimensionally reduced (i.e. two dimensional) theory \cite{sakai95}.
Moreover, in the $P_{\perp} = 0$ sector,
the mass squared operator $M^2$ is given by
$M^2=2P^+P^-$.

As we mentioned earlier, in order to render the bound state equations
numerically tractable, the transverse
momentum of partons must be truncated.
First, we introduce the Fourier expansion for the fields $\phi$ and $\psi$,
where the transverse space-time coordinate $x^{\perp}$ is periodically
identified:
\bea
\lefteqn{
\phi_{ij}(0,x^-,x_\perp) =} & & \nonumber \\
& &
\frac{1}{\sqrt{2\pi L}}\sum_{n^{\perp} = -\infty}^{\infty}
\int_0^\infty
 \frac{dk^+}{\sqrt{2k^+}}\left[
 a_{ij}(k^+,n^{\perp})e^{-{\rm i}k^+x^- -{\rm i}
\frac{2 \pi n^{\perp}}{L} x_\perp}+
 a^\dagger_{ji}(k^+,n^{\perp})e^{{\rm i}k^+x^- +
{\rm i}\frac{2 \pi n^{\perp}}{L}  x_\perp}\right]
\nonumber\\
\lefteqn{
\psi_{ij}(0,x^-,x_\perp) =} & & \nonumber \\
& & \frac{1}{2\sqrt{\pi L}}\sum_{n^{\perp}=-\infty}^{\infty}\int_0^\infty
 dk^+\left[b_{ij}(k^+,n^{\perp})e^{-{\rm i}k^+x^- -
{\rm i}\frac{2 \pi n^{\perp}}{L} x_\perp}+
 b^\dagger_{ji}(k^+,n^\perp)e^{{\rm i}k^+x^- +{\rm i}
\frac{2 \pi n^{\perp}}{L} x_\perp}\right]
\nonumber
\eea
Substituting these into the (anti)commutators (\ref{comm}),
one finds:
\begin{eqnarray}
\left[a_{ij}(p^+,n_\perp),a^\dagger_{lk}(q^+,m_\perp)\right]
&=&
\delta(p^+ -q^+)\delta_{n_\perp,m_\perp}
\left( \delta_{il}\delta_{jk} - \frac{1}{N}\delta_{ij} \delta_{lk} \right)
\\
\left\{b_{ij}(p^+,n_\perp),b^\dagger_{lk}(q^+,m_\perp)\right\}
&=&
\delta(p^+ -q^+)\delta_{n_\perp,m_\perp}
\left( \delta_{il}\delta_{jk} - \frac{1}{N}\delta_{ij} \delta_{lk} \right).
\end{eqnarray}
The supercharges now take the following form:
\bea\label{TruncSch}
&&Q^+={\rm i}2^{1/4}\sum_{n^{\perp}\in {\bf Z}}\int_0^\infty dk\sqrt{k}\left[
b_{ij}^\dagger(k,n^\perp) a_{ij}(k,n^\perp)-
a_{ij}^\dagger(k,n^\perp) b_{ij}(k,n^\perp)\right],\\
\label{Qminus}
&&Q^-=\frac{2^{7/4}\pi {\rm i}}{L}\sum_{n^{\perp}\in {\bf Z}}\int_0^\infty dk
\frac{n^{\perp}}{\sqrt{k}}\left[
a_{ij}^\dagger(k,n^\perp) b_{ij}(k,n^\perp)-
b_{ij}^\dagger(k,n^\perp) a_{ij}(k,n^\perp)\right]+\nonumber\\
&&+ {{\rm i} 2^{-1/4} {g} \over \sqrt{L\pi}}
\sum_{n^{\perp}_{i} \in {\bf Z}} \int_0^\infty dk_1dk_2dk_3
\delta(k_1+k_2-k_3) \delta_{n^\perp_1+n^\perp_2,n^\perp_3}
\left\{ \frac{}{} \right.\nonumber\\
&&{1 \over 2\sqrt{k_1 k_2}} {k_2-k_1 \over k_3}
[a_{ik}^\dagger(k_1,n^\perp_1) a_{kj}^\dagger(k_2,n^\perp_2)
b_{ij}(k_3,n^\perp_3)
-b_{ij}^\dagger(k_3,n^\perp_3)a_{ik}(k_1,n^\perp_1)
a_{kj}(k_2,n^\perp_2) ]\nonumber\\
&&{1 \over 2\sqrt{k_1 k_3}} {k_1+k_3 \over k_2}
[a_{ik}^\dagger(k_3,n^\perp_3) a_{kj}(k_1,n^\perp_1) b_{ij}(k_2,n^\perp_2)
-a_{ik}^\dagger(k_1,n^\perp_1) b_{kj}^\dagger(k_2,n^\perp_2)
a_{ij}(k_3,n^\perp_3) ]\nonumber\\
&&{1 \over 2\sqrt{k_2 k_3}} {k_2+k_3 \over k_1}
[b_{ik}^\dagger(k_1,n^\perp_1) a_{kj}^\dagger(k_2,n^\perp_2)
a_{ij}(k_3,n^\perp_3)
-a_{ij}^\dagger(k_3,n^\perp_3)b_{ik}(k_1) a_{kj}(k_2,n^\perp_2) ]\nonumber\\
&& ({ 1\over k_1}+{1 \over k_2}-{1\over k_3})
[b_{ik}^\dagger(k_1,n^\perp_1) b_{kj}^\dagger(k_2,n^\perp_2)
b_{ij}(k_3,n^\perp_3)
+b_{ij}^\dagger(k_3,n^\perp_3) b_{ik}(k_1,n^\perp_1) b_{kj}(k_2,n^\perp_2)]
 \left. \frac{}{}\right\}. \nonumber \\
\eea
We now perform the truncation procedure; namely,
in all sums over the transverse momenta $n^{\perp}_{i}$
appearing in the above expressions for the supercharges, we
restrict summation to the following allowed momentum
modes: $n^{\perp}_{i}=0,\pm 1$. More generally, the truncation procedure
may be defined by $|n^{\perp}_{i}|\le N_{max}$, where $N_{max}$ is
some positive integer. In this work, we consider the simple
case $N_{max}=1$.
Note that this prescription is
symmetric, in the sense that there are as many positive modes as
there are negative ones. In this way we
retain parity symmetry in the transverse
direction.

How does such a truncation affect the
supersymmetry properties of the theory?
Note first that an operator relation $[A,B]=C$ in the
full theory is not expected to hold in the truncated formulation.
However, if A is
quadratic in terms of fields (or in terms of creation and
annihilation operators),
one can show that the relation $[A,B]=C$ implies
$$
[A_{tr},B_{tr}]=C_{tr}
$$
for the truncated operators $A_{tr}$,$B_{tr}$, and $C_{tr}$.
In our case, $Q^+$ is quadratic, and so the relations
$\{Q_{tr}^+,Q_{tr}^+\}=2\sqrt{2}P_{tr}^+$ and $\{Q_{tr}^+,Q_{tr}^-\}=0$ are
true in the $P_\perp=0$ sector of the truncated theory.
The $\{Q_{tr}^-,Q_{tr}^-\}$
however is not equal to $2\sqrt{2}P_{tr}^-$. So the diagonalization of
$\{Q_{tr}^-,Q_{tr}^-\}$ will yield a different bound state spectrum
than the one obtained after diagonalizing $2\sqrt{2}P_{tr}^-$.
Of course the two spectra should agree in the limit
$N_{max}\rightarrow\infty$. At any finite truncation,
however, we have the liberty to
diagonalize any one of these operators.
This choice specifies our regularization scheme.

Choosing to diagonalize the light-cone
supercharge, however, has an important advantage:
{\em the spectrum is exactly supersymmetric for
any truncation}. In contrast, the spectrum of the Hamiltonian becomes
supersymmetric only in the $N_{max}\rightarrow\infty$
limit\footnote{If one chooses
anti-periodic boundary conditions in the $x^-$ coordinate
for fermions, then there is no choice; one can only
diagonalize the light-cone Hamiltonian.
See \cite{dak93} for more details on this approach.}.

To summarize, we have introduced a truncation procedure
that facilitates a numerical study of the bound state problem, and
preserves supersymmetry.
The interesting property of the light-cone supercharge $Q^-$
[Eqn(\ref{Qminus})] is the
presence of a gauge coupling constant as an independent variable,
which does not appear in the study of two dimensional theories.
In the next section, we will study how
variations in this coupling affects the bound states
in the theory.

\section{Numerical Results: Bound State Solutions}
\label{numerical}
In order to implement the DLCQ formulation of the bound state problem -- which
is tantamount to imposing periodic boundary conditions
$x^- = x^- + 2 \pi R$ \cite{yamawaki} -- we simply restrict the
light-cone momentum variables $k_i$ appearing in the expressions for $Q^+$
and $Q^-$ to the following discretized set of momenta:
$\left\{ \frac{1}{K} P^+, \frac{2}{K} P^+, \frac{3}{K} P^+,\dots,
\right\}$. Here, $P^+$ denotes the total light-cone momentum of a
state, and may be thought of as a fixed constant, since it is
easy to form a Fock basis that is already diagonal with respect to the
operator $P^+$ \cite{pb85}. The integer $K$ is called the `harmonic
resolution', and $1/K$ measures the coarseness of our discretization.
The continuum limit is then recovered by taking the limit
$K \rightarrow \infty$. Physically, $1/K$ represents the smallest
positive unit of longitudinal momentum-fraction allowed for each
parton in a Fock state.

\begin{figure}[h]
\begin{center}
\epsfig{file=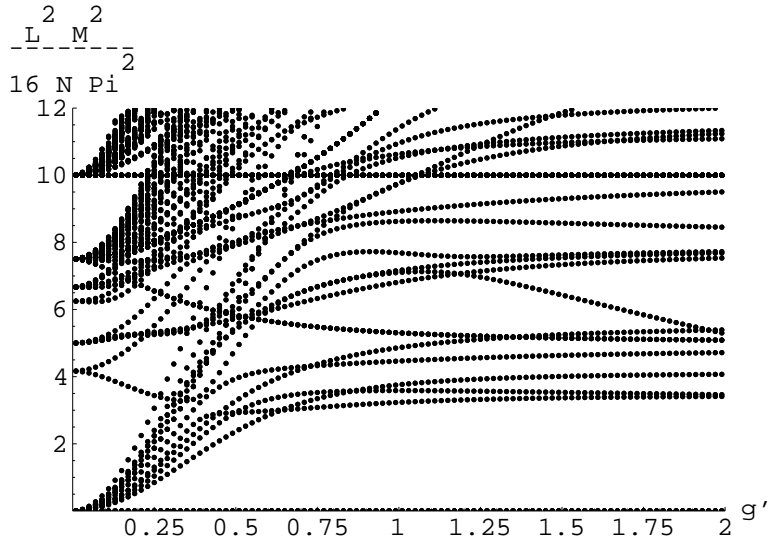}
\end{center}
\caption{{\small Plot of bound state mass squared $M^2$
in units $16 \pi^2 N/ L^2$
as a function of the dimensionless coupling $0 \leq g' \leq 2$,
defined by $(g')^2 = g^2 N L/16 \pi^3$,
at $N=1000$ and $K=5$. Boson and fermion masses are identical.
} \label{mass1}}
\end{figure}

Of course, as soon as we implement the DLCQ procedure, which is
specified unambiguously by the harmonic resolution $K$,
and cut-off transverse momentum modes via the constraint
$|n_i^{\perp}| \leq N_{max}$,
the integrals
appearing in the definitions for $Q^+$ and $Q^-$ are replaced by finite sums,
and so the eigen-equation $2P^+P^-|\Psi\rangle = M^2 |\Psi\rangle$
is reduced to a finite matrix diagonalization problem.
In this last step we use the fact that $P^-$ is proportional
to the square of the light-cone supercharge\footnote{
Strictly speaking, $P^- = \frac{1}{\sqrt{2}}(Q^-)^2$ is an identity
in the continuum theory, and a {\em definition} in the compactified
theory, corresponding to the SDLCQ prescription \cite{sakai95,alp99}.}
$Q^-$. In the present work, we are able to perform numerical
diagonalizations for $K=2,3,4$ and $5$ with the help of Mathematica and a
desktop PC.
\begin{figure}[ht]
\begin{center}
\epsfig{file=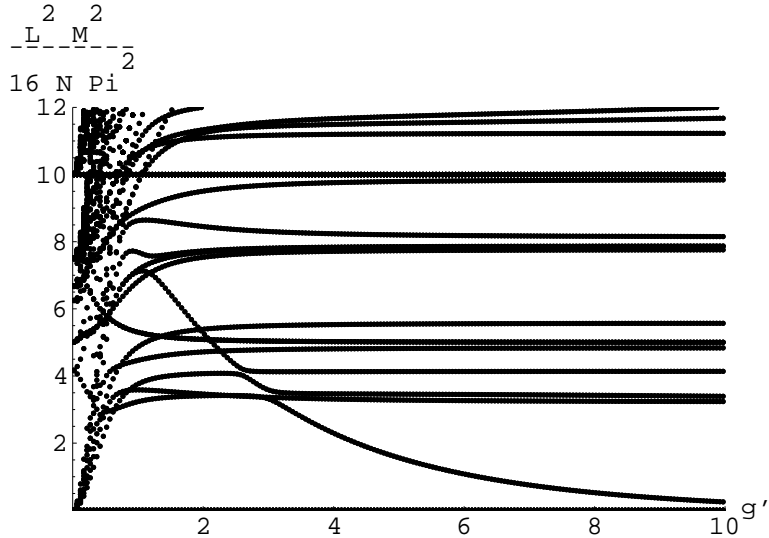}
\end{center}
\caption{{\small Plot of bound state mass squared $M^2$
in units $16 \pi^2 N/ L^2$
as a function of the dimensionless coupling $0 \leq g' \leq 10$,
defined by $(g')^2 = g^2 N L/16 \pi^3$,
at $N=1000$ and $K=5$. Note the appearance of a new massless
state at strong coupling.} \label{mass2}}
\end{figure}
In Figure \ref{mass1}, we plot the bound state mass squared
$M^2$, in units $16 \pi^2 N/L^2$,
as a function of the dimensionless coupling
$g' = g \sqrt{N L}/4 \pi^{3/2}$, in the range $0 \leq g' \leq 2$.
We consider the specific case $N=1000$, although our algorithm
can calculate masses for any choice of
$N$, since it enters our calculations as an
algebraic variable.
Since there is an exact boson-fermion mass degeneracy,
one needs only diagonalize the mass matrix $M^2$
corresponding to the bosons. For $K=5$, there are precisely
600 bosons and 600 fermions in the truncated light-cone Fock space,
so the mass matrix that needs to be diagonalized has dimensions
$600 \times 600$. At $K=4$, there are $92$ bosons and $92$ fermions,
while at $K=3$, one finds $16$ bosons and $16$ fermions.

In Figure \ref{mass2}, we plot the bound state spectrum in
the range $0 \leq g' \leq 10$. It is apparent now that
new massless states appear in the strong coupling limit
$g' \rightarrow \infty$.
\begin{figure}[h]
\begin{center}
\epsfig{file=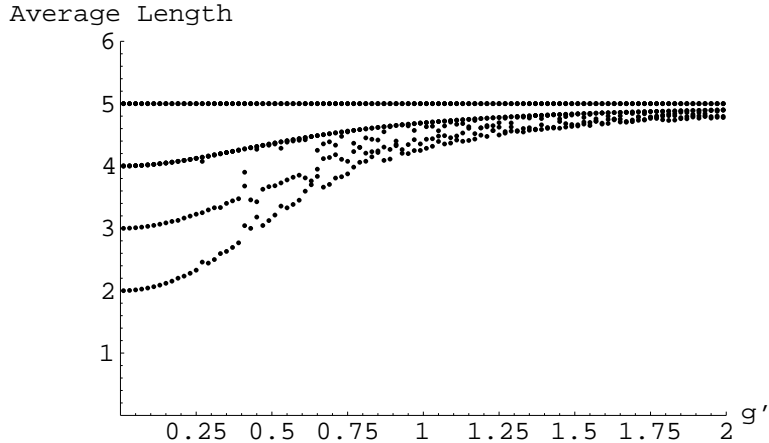}
\end{center}
\caption{{\small Plot of average length for the eight massless
bosonic states
as a function of the dimensionless coupling $g'$,
defined by $(g')^2 = g^2 N L/16 \pi^3$,
at $N=1000$ and $K=5$. Note that the states attain the maximum possible
length allowed by the resolution $K=5$
in the limit of strong coupling.} \label{lengths}}
\end{figure}

An interesting property of the spectrum is the
presence of exactly massless states
that persist for all values of the coupling $g'$.
For $K=5$, there are 16 such states (8 bosons and 8 fermions).
At $K=4$, one finds 8 states (4 bosons and 4 fermions)
that are exactly massless
for any coupling, while for $K=3$, there are 4 states
(two bosons and two fermions) with
this property.
We will have more to say regarding these states in the next section,
but here we note that the structure of these states become
`string-like' in the strong coupling limit. This is illustrated
in Figure \ref{lengths}, where we plot the `average length'
(or average number of partons)
of each of these massless states\footnote{The `noisiness' in this
plot for larger values of $g'$ reflects the ambiguity of choosing
a basis for the eigen-space, due to the exact mass degeneracy
of the massless states.}.
This quantity is obtained
by counting the  number of partons in each  Fock state
that comprises a massless bound state, appropriately
weighted by the modulus of the wave function squared.
Clearly, at strong coupling, the average number of partons saturates
the maximum possible value allowed by the resolution -- in this
case 5 partons. The same behavior is observed at lower resolutions.
Thus, in the continuum limit $K \rightarrow \infty$, we expect
the massless states in this theory to become string-like
at strong coupling.
\begin{table}[h]
\begin{center}
\begin{tabular}{|c|c|c|c|c|}
\hline
\multicolumn{5}{|c|}{Comparison Between $1+1$ and $2+1$ Spectra} \\
\hline
- &
$1+1$ Model &
\multicolumn{3}{|c|}{Rescaled $2+1$ Model}
 \\
\hline
$K$ & - & $g'=.01$ & $g'=.1$ & $g'=1.0$ \\
\hline
$K=5$ & 15.63 & 15.5 & 15.17 & 3.7 \\
 & 18.23 & 17.6 & 17.9 & 3.5 \\
 & 21.8 & 21.3 & 21.7 & 3.2 \\
\hline
$K=4$ & - & - & - & - \\
 & 18.0 & 17.99 & 17.6 & 3.56 \\
 & 21.3 & 21.3 & 21.0 & 3.1 \\
\hline
$K=3$ & - & - & - & - \\
  & - & - & - & - \\
 & 20.2 & 20.2 & 19.8 & 3.1 \\
\hline
\end{tabular}
\caption{{\small Values for the mass squared $M^2$, in units
${\tilde g}^2 N/\pi$, with ${\tilde g}^2 = g^2/L$,
for bound states in the dimensionally
reduced ${\cal N}=(1,1)$ model, and
the $2+1$ model studied here.
The quantity ${\tilde g}$ is identified as the gauge
coupling in the $1+1$ model. We set $K=3,4$ and $5$, and $N=1000$.
Note that the comparison of masses between
the $1+1$ model, and the (re-scaled) $2+1$ model
is good only at weak coupling $g'$.}
\label{masspredictions}}
\end{center}
\end{table}

One interesting property of the model studied here is
the manifest ${\cal N}=(1,1)$
supersymmetry in the $P^{\perp}=0$ momentum sector, by virtue
of the supersymmetry relations (\ref{superr}).
Moreover, if we consider retaining only the zero mode $n_i^{\perp}=0$,
then the light-cone supercharge $Q^-$ for the $2+1$ model is identical
to the $1+1$ dimensional ${\cal N}=(1,1)$ supersymmetric
Yang-Mills theory studied in \cite{sakai95,alp98a,alp98b}, after a rescaling
by the factor $1/g'$.
(This is equivalent to expressing the mass squared
$M^2$ in units ${\tilde g}^2 N/\pi$, where ${\tilde g}=g/\sqrt{L}$.
The quantity ${\tilde g}$ is then identified as the gauge coupling
in the $1+1$ theory.)
 We may therefore think of the
additional transverse degrees of freedom in the $2+1$
model, represented by the modes $n^{\perp} = \pm 1$, as
a modification of the $1+1$ model. A natural question
that follows from this viewpoint is: How well does the
$1+1$ spectrum approximate the $2+1$ spectrum after
performing this rescaling? Before discussing the numerical
results summarized in Table \ref{masspredictions},
let us first attempt to predict what will happen
at small coupling $g'$. In this case, the coefficients
of terms in the rescaled Hamiltonian $P^-$ that
correspond to summing the
transverse momentum squared $|k^{\perp}|^2$ of partons in a state
will be large. So the low energy sector will be dominated
by states with $n^{\perp}=0$. i.e. those states that appear in the
Fock space of the ${\cal N}=(1,1)$ model in $1+1$ dimensions.
This is indeed supported by the results in Table \ref{masspredictions}.

For large coupling $g'$, however, it is clear that the
approximation breaks down. In fact, one can show that
the tabulated masses in the rescaled $2+1$ model
tend to zero in the strong
coupling limit, which  eliminates any scope for making
comparisons between the two and three dimensional models.

Thus, the non-perturbative problem of solving dimensionally
reduced models in $1+1$ dimensions can only provide information
about bound state masses in the corresponding {\em weakly
coupled} higher
dimensional theory.
\section{Analytical Results: The Massless Sector}
\label{analytical}

In the previous section we presented the results of studying the bound state
problem using numerical methods. In performing such a study we
conveniently chose the simplest nontrivial truncation
of the transverse momentum modes; namely,  $n_\perp=0,\pm 1$.
Surprisingly, such a simple scheme provided many interesting
insights concerning the massless and massive sector. In particular
we see that there are three types of massless states; those that are massless
only at $g=0$ or $g=\infty$ (but not both), and those
that are massless for any value of the coupling. In this section, we will
analyze only the massless sector of the theory, and show that the observed
properties of the
spectrum  with the truncation $n^{\perp}=0,\pm 1$ also persists
if we include higher modes: $n^{\perp} =0,\pm 1,\pm2,\dots,\pm N_{max}$.
We therefore consider
the model with supercharges given by (\ref{TruncSch}) and
(\ref{Qminus}),
and restrict summation of transverse momentum modes
via the constraint $|n^{\perp}|\leq N_{max}$.

For states carrying positive light-cone momentum,
$P^+$ is never zero, and so massless states must satisfy
the equation $P^-|\Psi \rangle = 0$, which, using
the relation $P^-=\frac{1}{\sqrt{2}}(Q^-)^2$, and hermiticity of $Q^-$,
reduces to
\be\label{msls}
Q^-|\Psi\rangle=0.
\ee
This is the equation we wish to study in detail.

We begin with an analysis of the weak coupling limit of the theory.
This limit means that the dimensionless coupling constant
is small: i.e. $g \sqrt{L} \ll 1$.
We will
consider the strong--weak coupling behavior of the theory on a cylinder with
fixed circumference $L$ so it is convenient to choose the units in which
$L=1$ for this discussion.
The supercharge (\ref{Qminus}) consists of two
parts: one is proportional to the coupling and the other is
coupling--independent:
\be\label{QminG}
Q^-=Q_{\perp}+g{\tilde Q}.
\ee
So at $g=0$, the equation (\ref{msls}) reduces to
$Q_{\perp}|\Psi\rangle=0$,
which means that $|\Psi\rangle$ may be viewed as a state
in the Fock space of the two dimensional
${\cal N}=(1,1)$ super Yang-Mills theory,
which may be obtained by dimensional reduction of the $2+1$ theory.
Thus the massless states at $g=0$ are states with any combination of
$a^\dagger(k,0)$ and $b^\dagger(k,0)$ modes, and no partons with nonzero
transverse
momentum.

What happens with these massless states when one switches on
the coupling? To
answer this question, we need some information about the
behavior of states
as functions of the coupling. We assume that wave functions are analytic
in terms of $g$ at least in the vicinity of $g=0$.
This means that in this region any
massless state $|\Psi\rangle$ may be written in the form:
\be
|\Psi\rangle=\sum_{n=0}^{\infty} g^n |n\rangle,
\ee
where states $|n\rangle$ are coupling independent.
Then using relation (\ref{QminG}),
the $g$--dependent equation (\ref{msls}) may be written as an
infinite system of
relations between different $|n\rangle$:
\bea
Q_\perp |0\rangle=0,\\
\label{InfSyst}
Q_\perp |n\rangle+{\tilde Q}|n-1\rangle=0, \qquad n>0.
\eea
The first of these equations was already used to exclude partons
carrying non-zero transverse momentum, which is a property
of the massless bound states at zero coupling.
The second equation is non-trivial. Let us
consider two different subspaces in the theory.
The first of these subspaces consists of
states with no creation operators for transverse modes which we will label
$1$. The other is the
complement of this space in which the operator $Q_\perp$ is invertible and
we label this space 2.
Equation (\ref{InfSyst}) defines the recurrence relation when ${\tilde Q}
|n-1\rangle$ is in subspace $2$:
\be\label{recur}
|n\rangle=-Q_\perp^{-1}\left(\left.{\tilde Q}
|n-1\rangle\right|_2\right),
\ee
The consistency condition is that  projection of ${\tilde Q}
|n-1\rangle$ in subspace $1$ is zero,
\be\label{consist}
\left.{\tilde Q}|n-1\rangle\right|_1=0.
\ee
This condition implies that not all states of the two dimension theory, $g=0$ ,
may be extended to such states at arbitrary $g$ using (\ref{recur}).
Taking $n=1$, (\ref{consist}) implies that $|0\rangle$
is a massless state of the  dimensionally reduced theory.  The numerical
solutions, of course, show
this correspondence between the  $2+1$ and
$1+1 $\cite{sakai95,alp98a,alp98b} massless bound states.  Starting from a
massless state of the two dimensional theory, and we construct states
$|n\rangle$ using (\ref{recur}), and for which (\ref{consist}) is always
satisfied.
Then $|\Psi\rangle$
may be found from summing a geometric series:
\be\label{3dmslsst}
|\Psi\rangle=\sum_{n=0}^\infty (-gQ_\perp^{-1}{\tilde Q})^n|0\rangle=
\frac{1}{1+gQ_\perp^{-1}{\tilde Q}}|0\rangle.
\ee
So, starting from the massless state of the two dimensional
${\cal N}=(1,1)$ model, one can always
construct unique massless states in the three dimensional theory
at least in the vicinity of $g=0$.

The state (\ref{3dmslsst}) turns out to be massless for any value of the
coupling:
\be
Q^-|\Psi\rangle=Q_{\perp}(1+gQ_\perp^{-1}{\tilde Q})
  \frac{1}{1+gQ_\perp^{-1}{\tilde Q}}|0\rangle=Q_{\perp}|0\rangle=0,
\ee
though the state itself is  dependent on g.
Thus, we have shown that massless states of the three
dimensional theory, at nonzero
coupling, can be constructed from massless states of
the corresponding model in two
dimensions. All other states containing only two dimensional modes can also be
extended to the eigenstates of the full theory. But such eigenstates are
massless only at zero coupling. Assuming analyticity, one can then
show that their masses grow linearly at $g$ in the vicinity of zero.
Such behavior also agrees with our numerical results.

To illustrate the general construction explained above we consider one simple
example. Working in DLCQ at resolution $K=3$ we choose a special
two dimensional massless state\footnote{
The state $|0 \rangle$ denotes a massless state,
while $|vac \rangle$ represents the light-cone vacuum.} 
\cite{sakai95,alp98a,alp98b}:
\be
|0\rangle=\mbox{tr} (a^\dagger (1,0)a^\dagger (2,0))|vac\rangle.
\ee
Then in the SU($N$) theory we find:
\bea
{\tilde Q}|0\rangle&=&\frac{3}{2\sqrt{2}}\left[\mbox{tr}\left(a^\dagger (1,0)
(b^\dagger (1,-1)a^\dagger (1,1)-a^\dagger (1,1)b^\dagger (1,-1)+\right.\right.
\nonumber\\
&+&\left.\left.b^\dagger (1,1)a^\dagger (1,-1)-a^\dagger (1,-1)b^\dagger (1,1))
\right)
\right]|vac\rangle,\\
|1\rangle&=&-Q_\perp^{-1}{\tilde Q}|0>=
-\frac{\sqrt{L}}{4\pi^{3/2}}
\frac{3}{2\sqrt{2}}
\left(a^\dagger (1,0)a^\dagger (1,-1)a^\dagger (1,1)-\right.\nonumber\\
&-&\left. a^\dagger (1,0)a^\dagger (1,1)a^\dagger (1,-1)\right)
|vac\rangle \\
{\tilde Q}|1\rangle&=&0.
\eea
The last equation provides the consistency condition (\ref{consist}) for
$n=2$, and it also shows that for this special example we have only
two states
$|0\rangle$ and $|1\rangle$,
instead of a general infinite set.
The matrix form of the
operator $1+gQ_\perp^{-1}{\tilde Q}$ in the $|0\rangle , |1\rangle$
basis is
\be
1+gQ_\perp^{-1}{\tilde Q}=
\left(\begin{array}{cc}
1&-g\\0&1\end{array}\right)=
\left(\begin{array}{cc}
1&g\\0&1\end{array}\right)^{-1}.
\ee
Then the solution of (\ref{3dmslsst}) is
\bea
&&|\Psi\rangle=|0\rangle+g|1\rangle=
\mbox{tr} (a^\dagger (1,0)a^\dagger (2,0))|vac\rangle+\\
&&+\frac{g\sqrt{L}}{4\pi^{3/2}}\frac{3}{2\sqrt{2}}
\left(a^\dagger (1,0)a^\dagger (1,1)a^\dagger (1,-1)-
a^\dagger (1,0)a^\dagger (1,-1)a^\dagger (1,1)\right)|vac\rangle.
\nonumber
\eea
This state was observed numerically, and the dependence of the wave
function on the coupling constant is precisely the
one given by the last formula.

In principle, we can determine the wave functions of all massless states using
this formalism. Our procedure has an important advantage over a direct
diagonalization of the three dimensional supercharge. Firstly,
in order to find two dimensional massless states,
one needs to diagonalize the corresponding supercharge
\cite{sakai95}.
However, the dimension of the relevant Fock space is
much less than the three dimensional theory
(at large resolution $K$, the ratio of these
dimensions is of order $(N_{max}+1)^{\alpha K}$, $\alpha\sim 1/4$). The
extension of the two dimensional massless solution
into a massless solution of the three dimensional theory
requires diagonalizing a matrix which has a smaller
dimension than the original problem in three dimensions.
Thus, if one is only interested in the massless sector
of the three dimensional theory, the most efficient
way to proceed in DLCQ calculations is to
solve the two dimensional theory, and then to
upgrade the massless states to massless solutions in three dimensions.

Finally, we will make some comments on
bound states at very strong coupling. Of
course, we have states (\ref{3dmslsst}) which are massless at any coupling,
but our numerical calculation show there are additional states which become
massless at $g=\infty$ (see Figure \ref{mass2}).
To discuss these state it is convenient to consider
\be
{\bar Q}^-=\frac{1}{g}Q_{\perp}+{\tilde Q}
\ee
instead of $Q^-$, and perform the strong coupling expansion. Since we are
interested only in massless states, the absolute
normalization doesn't matter.  We
repeat all the arguments used in the weak coupling case: first,
we introduce the space $1^*$  where ${\tilde Q}$ can not be inverted,
and its orthogonal complement $2^*$.
Then any state from $1^*$ is massless at $g=\infty$, but assuming the
expansion
\be
|\Psi\rangle=\sum_{n=0}^{\infty} \frac{1}{g^n} |n\rangle^*
\ee
at large enough $g$, one finds the analogs of (\ref{recur}) and
(\ref{consist}):
\bea
|n\rangle^*=-{\tilde Q}^{-1}\left(\left.Q_\perp|n-1\rangle^*
\right|_{2^*}\right),\\
\left.Q_\perp|n-1\rangle^*\right|_{1^*}=0.
\eea
As in the small coupling case, there are two possibilities: either
one can construct all states $|n\rangle^*$
satisfying the consistency conditions, or
at least one of these conditions fails.
The former case corresponds to the massless state in the vicinity
of $g=\infty$, which can be extended to the
massless states at all couplings.
The states constructed in this way -- and ones given by
(\ref{3dmslsst}) -- define the same subspace.
In the latter case,
the state is massless at $g=\infty$, but it acquires
a mass at finite coupling.
There is a big difference,
however, between the weak and strong coupling cases.
While the kernel of $Q_\perp$
consists of "two dimensional" states,
the description of the states annihilated by
${\tilde Q}$ is a nontrivial dynamical problem. Since the massless states
can be constructed starting from either $g=0$ or $g=\infty$,
we don't have to solve this problem to build them.
If, however, one wishes to show that
massless states become long in the strong coupling limit (there is numerical
evidence for such behavior -- see Figure \ref{lengths}),
the structure of $1^*$ space becomes important,
and we leave this question for future investigation.

\section{Discussion}
\label{summary}
In this work, we considered the bound states of
three dimensional SU($N$) ${\cal N}=1$ super-Yang-Mills
defined on the compactified space-time ${\bf R} \times S^1 \times S^1$.
In particular, we compactified the light-cone coordinate
$x^-$ on a light-like circle via DLCQ, and wrapped the remaining
transverse coordinate $x^{\perp}$ on a spatial circle.
We showed explicitly that the supersymmetric form of DLCQ
(or `SDLCQ') employed in recent studies of two dimensional supersymmetric
gauge theories extends naturally in $2+1$ dimensions,
which resulted in an exactly supersymmetric spectrum.
We also showed that the ${\cal N}=1$
supersymmetry is enhanced to 
${\cal N}=(1,1)$ in a reference frame with vanishing
total transverse momentum $P^{\perp}=0$. 
The supersymmetric theory considered here is actually 
super-renormalizable\footnote{Ultraviolet renormalization
is never an issue here, since we have truncated the transverse momentum
modes, which acts as a regulator. DLCQ regulates any longitudinal
divergences for vanishing $k^+$.}.

By retaining only the first few excited modes in the transverse direction,
we were able to solve for bound state wave functions and masses
numerically by diagonalizing the  discretized light-cone supercharge. The
results of our numerical calculations for bound state masses are
summarized in Figures
\ref{mass1} and \ref{mass2}. The theory exhibits a stable spectrum at
both small and large coupling.  
In Table \ref{masspredictions},
we compared solutions of the $2+1$
dimensional theory to corresponding solutions in the dimensionally
reduced $1+1$ theory, after an appropriate rescaling of
the $1+1$ dimensional
coupling constant, and observed that the lower dimensional theory provides
a good approximation to the low energy spectrum of
the higher dimensional theory at weak coupling only.
Any scope for making comparisons breaks down, however, at intermediate
and strong coupling.
One also notes a smooth dependence of bound state
masses in terms of the DLCQ harmonic resolution $K$, 
a fact that was observed previously in studies of
related supersymmetric models
\cite{alp98b}. 

Interestingly, we  find exactly massless states that persist
for all values of the gauge coupling. 
These states should be contrasted with those that are massless
only at the extreme values
$g=0$, or $g=\infty$ (but not both).  

An analytical treatment of the massless states
revealed a connection with the massless solutions
of the corresponding dimensionally reduced model
\cite{sakai95,alp98a,alp98b}. We have shown that the wave functions of the
massless states that remain massless for all values
of the gauge coupling are in one-to-one
correspondence with the massless states of the $1+1$ dimensional theory.
This fact seems to be non-trivial, since the concept of mass is 
defined differently in the two theories. 
From the three dimensional point of view, the states in
the Hilbert space that are naturally associated with the two dimensional
Fock space (i.e. those states made up from partons with zero transverse
momentum) are massless at $g=0$. 

The bound state structure of the massless states in
the $2+1$ theory were also studied for different couplings, and summarized
in Figure \ref{lengths}, where we plotted the average number
of partons for each of the massless solutions.
 We concluded that in the decompactified limit $K \rightarrow \infty$, these
massless states must become string-like in the strong coupling
limit.

Evidently, it would be interesting to relate these
observations with the recent claim that strongly-coupled
super-Yang-Mills theory corresponds to string theory in an
anti-de Sitter background \cite{maldacena}.
Of course, the techniques we have employed in this study may be
applied to any supersymmetric
gauge theory defined on a suitably compactified space-time.
This should facilitate a more general study of the strongly
coupled dynamics of super-Yang-Mills theories, and in
particular, allow one to scrutinize more carefully 
the string-like properties
of Yang-Mills theories.




\vfil

\end{document}